\documentclass[12pt,preprint]{aastex}

\def\lea{\mathrel{<\kern-1.0em\lower0.9ex\hbox{$\sim$}}}
\def\gea{\mathrel{>\kern-1.0em\lower0.9ex\hbox{$\sim$}}}

\shorttitle{ACS Survey of Galactic Globular Clusters}
\shortauthors{Sarajedini et al.}

\begin{document}


\title{The ACS Survey of Galactic Globular Clusters. I.\thanks{
Based on observations with the NASA/ESA {\it Hubble Space Telescope},
obtained at the Space Telescope Science Institute, which is operated
by AURA, Inc., under NASA contract NAS 5-26555, under programs
GO-10775 (PI: Sarajedini).
}
 \\ Overview and Clusters Without Previous {\sl HST} Photometry}

\author{Ata Sarajedini}
\affil{Department of Astronomy, University of Florida, 211 Bryant Space Science
Center, Gainesville, FL 32611
\email{ata@astro.ufl.edu} }

\author{Luigi R. Bedin}
\affil{European Southern Observatory, Garching,
            Karl-Schwarzschild-Str.\ 2, D-85748, D, EU
            \email{lbedin@eso.org}}

\author{Brian Chaboyer and Aaron Dotter}
\affil{Department of Physics and Astronomy, Dartmouth College, 
6127 Wilder Laboratory, Hanover, NH 03755 \email{chaboyer@heather.dartmouth.edu,
Aaron.L.Dotter@dartmouth.edu} }

\author{Michael Siegel}
\affil{University of Texas, McDonal Observatory, 1 University Station, C1402, Austin TX, 78712
\email{siegel@astro.as.utexas.edu} }

\author{Jay Anderson}
\affil{Department of Physics and Astronomy, Rice University MS-108, Houston, TX 77005
\email{jay@eeyore.rice.edu} }

\author{Antonio Aparicio}
\affil{University of La Laguna and Instituto de Astrof\'\i sica de Canarias
La Laguna, Canary Islands, Spain
\email{antapaj@iac.es} }


\author{Ivan King}
\affil{Dept. of Astronomy, Univ. of Washington, Box 351580, Seattle, WA 98195-1580
\email{king@astro.washington.edu} }

\author{Steven Majewski}
\affil{Dept. of Astronomy, University of Virginia,
P.O. Box 400325, Charlottesville, VA 22904-4325
\email{srm4n@virginia.edu} }

\author{A. Mar\' \i n-Franch}
\affil{Department of Astronomy, University of Florida, 211 Bryant Space Science
Center, Gainesville, FL 32611
\email{amarin@astro.ufl.edu} }



\author{Giampaolo Piotto}
\affil{Dipartimento di Astronomia, Universit\`{a} di Padova, 35122 Padova, Italy 
\email{piotto@pd.astro.it} }

\author{I. Neill Reid}
\affil{Space Telescope Science Institute, 3700 San Martin Drive, Baltimore MD 21218
\email{inr@stsci.edu} }

\author{Alfred Rosenberg}
\affil{Instituto de Astrof\'\i sica de Canarias, V\'\i a L\'{a}ctea s/n, E-38200 La Laguna, 
Spain
\email{alf@iac.es} }



\begin{abstract}
We present the first results of a large ACS Survey of Galactic globular clusters. This
Hubble Space Telescope ({\sl HST}) Treasury project is designed to obtain photometry 
with S/N $\gea$ 10 for main sequence stars with masses $\gea$ 0.2$M_{\odot}$
in a sample of globulars using the Advanced Camera for Surveys (ACS)
Wide Field Channel. Here we focus on clusters
without previous {\sl HST} imaging data. These include 
NGC~5466, 6779, 5053, 6144, Palomar~2, E~3, Lyng\aa~7, Palomar~1, and
NGC~6366. Our CMDs extend reliably from the horizontal branch 
to as much as seven magnitudes fainter than the main sequence turnoff and
represent the deepest CMDs published to-date for these clusters. 
Using fiducial sequences for three standard clusters (M92, NGC~6752, and 47 Tuc) 
with well-known metallicities and distances, we perform main sequence fitting
on the target clusters in order to obtain estimates of their distances and reddenings. 
These comparisons along with fitting the cluster main sequences to theoretical
isochrones yield ages for the target clusters. We find that the majority
of the clusters have ages that are consistent with the standard clusters
at their metallicities. The exceptions are E~3 which appears $\sim$2 Gyr younger 
than 47 Tuc, and Pal 1, which could be as much as 8 Gyr younger than 47 Tuc.
\end{abstract}




\section{Introduction}

We present the first results of a 134-orbit Hubble Space Telescope ({\sl HST}) 
Treasury program to conduct an ACS Survey of Galactic Globular Clusters (GC). 
%
%
This project is designed to obtain uniform photometry with S/N $\gea$ 10 for stars
as faint as 0.2${\cal M}_{\odot}$ ($M_V$$\lea$10.7) along the main sequence for 
approximately half of the nearest known Milky Way GCs using the Wide Field 
Channel (WFC) of the Advanced Camera for Surveys (ACS). 
The survey will produce an image atlas 
%
%
and source catalog with astrometry and photometry for stars
in the target clusters using both newly obtained ACS observations as well as archival
ACS and Wide Field Planetary Camera 2 imaging, where available. In the spirit of the 
{\sl HST} Treasury concept, the overall goal of this ``legacy" survey is to investigate 
fundamental aspects of Galactic GCs (e.g., luminosity functions, 
reddenings, distances, ages, proper motions, binary fractions, to name a few) and 
provide a lasting contribution to cluster studies by creating a uniquely deep and uniform 
database of a large sample of Galactic GCs. 

Our target list includes 66 GCs chosen by a number of criteria, the most
important of which are proximity to the Sun ($(m-M)_0 \leq 16.5$) and 
low-reddening ($E(B-V) \leq 0.35$). However, we have also included a few
clusters of intrinsic interest such as those believed to be associated with the
Sagittarius dwarf spheroidal galaxy and a number of clusters in the direction
of the Galactic bulge. The full details of our target list will be
published in a forthcoming paper. 

This first paper in our survey is concerned with GCs in our target list
that have not been previously imaged by {\sl HST}. These include the metal-poor
clusters NGC~5466, 6779 (M56), 5053, and 6144, the intermediate
metallicity cluster Palomar~2, and the
metal-rich clusters E~3, Lyng\aa~7, Palomar~1, and NGC~6366. The clusters are
ordered, here as elsewhere in this paper, by metallicity. In the next section, 
we summarize the observations and data reduction, which are generally similar for 
all clusters in our program. The resultant color-magnitude diagrams
are presented and discussed in Sec.\ 3. Main sequence fits and comparisons 
with theoretical isochrones are included in Sec.\ 4 and 5, respectively.
Section 6 presents a summary of our results.


\section{Observations and Data Reduction}

The observations of our program clusters were obtained with the {\sl HST/ACS/WFC}
instrument in the F606W ($\sim$$V$) and F814W ($\sim$$I$) filters. Each cluster
was centered in the ACS field and observed for two orbits, one orbit for each filter, 
with one short exposure per filter (except for Pal~2) and four to five long exposure frames. 
The long exposures were dithered to fill in the gap between the two CCDs of ACS. 
Table 1 presents the log of the observations. 

The process of deriving photometry from the short and long exposures and
combining the results will be fully described in an upcoming paper 
(Anderson et al.\ 2007, in preparation). To summarize, 
we reduced each {\tt \_FLT} exposure independently using the 
program {\tt img2xym\_WFC.09x10}, which is documented in 
Anderson \& King (2006).  The program 
uses an array of 9$\times$10 PSFs to treat the spatial variability 
of the WFC PSF.  The routine also allows for a spatially constant
perturbation adjustment for the PSF to better match each individual
exposure.  (The perturbation adjustment for the PSF was possible 
only on exposures that had a sufficient number of well-exposed stars; 
this ruled out all of the short exposures.)

The above computer program went through each image pixel by pixel 
and used the PSF model to get a flux and a position for every 
source that had no brighter neighbors within 3 pixels and had at 
least 50 e$^-$'s above sky in a 3$\times$3-pixel aperture.  We correct 
the source's position for distortion using the prescription in 
Anderson (2006) and find the transformation from each frame into
a single reference frame centered on the cluster core. 

We then collate the deep star lists in the reference frame,
identifying a star wherever there are three or more coincident 
detections in the deep exposures
for each filter.
This collation procedure naturally removes any
cosmic rays or warm pixels that may have strayed into the
deep lists
for the individual exposures.  We now have one star list for each of
the two filters.  This list contains only stars that are below the
saturation limit in the deep exposures.  Since each star in this
list has been observed in a minimum of three exposures independently,
we also have an estimate of the error in its photometry from the
rms magnitude about the mean of the independent observations.

We supplement this list with the star lists for the short exposures.
We first find an empirical photometric-zeropoint offset between the 
short and deep exposures and add to the deep list for each filter 
any star in the short exposures that is within 0.75 magnitudes of 
saturation or brighter in the deep list.  Stars that are saturated in the short 
exposures are measured by fitting the PSF to their unsaturated pixels 
and are included in the list, even though their photometry and 
astrometry are less accurate.
Thus, in the end, we have a list of stars found in each of the two filters.  
Our final star catalog is produced by merging the 
F606W and F814W lists, keeping only stars found in both filters.  
It is important to emphasize at this point that this `first pass' photometry 
includes only the uncrowded stars (i.e. those with no brighter neighbors within 3 pixels).
The `final pass' reduction, where we relax this crowding restriction,
will include many more stars and will therefore be more complete. The
photometric catalog that the Treasury project makes available to the community
will include only the final pass data.

The resultant instrumental magnitudes must be corrected for the effects
of charge transfer efficiency (CTE) before being transformed to a standard
system. To facilitate this, we rely upon the formalism of Reiss \& Mack
(2004, hereafter RM2004). We begin by noting that all of our long exposure 
observations, except for those of E3, exhibit background sky values greater 
than $\sim$30 electrons.
Equation 2 of RM2004 suggests that, at these sky levels, the CTE correction is less
than $\sim$0.005 mag for all stellar magnitudes and y-coordinates. Even
in the case of E3 where the mean sky level is $\sim$10 electrons, Fig.\ 2 of 
RM2004 shows that the CTE correction is typically 0.01 mag. Based on this,
we choose not to correct the long exposure photometry for the CTE effect and
proceed to correct the short exposure photometry using 
Equation 2 of RM2004. These corrections are relatively small amounting
to $\sim$0.03 mag for the faintest stars on the short exposure frames roughly
equalling the expected random errors for these stars. To check 
the efficacy of these 
corrections, we examine the difference in magnitude of a given star between 
the long and short  exposures as a function of Y-position to be sure that there
is no trend present. Since the corrections are approximately the same
on the F606W and F814W filters, the colors are minimally affected. 

We calibrate the photometry to the ACS/WFC VEGAmag system following
the procedure given in Bedin et al.\ (2005) using the most
updated encircled energy distributions and the official zeropoints given by
Sirianni et al.\ (2005).
%
%
The result of this procedure is photometry that reliably extends from 
the horizontal branch (HB) to several magnitudes below the main sequence
turnoff (MSTO).  \footnote{Electronic versions of the photometry tables and fiducial
sequences are available at http://www.astro.ufl.edu/$\sim$ata/GC\_Treasury/Web\_Page/.}

\section{Color-Magnitude Diagrams}

The CMDs derived from our {\sl HST/ACS} observations of the target clusters
are shown in Figs. 1 through 9. In each case, the crosses indicate stars affected
by at least one saturated pixel in either or both of the F606W and F814W short
exposure images. All four of the metal-poor clusters display
predominantly blue HBs along with significant blue straggler sequences. 
Six of the nine clusters also show rich populations of unresolved
main sequence binaries as evidenced by the parallel sequences located
as much as 0.75 mag brighter than the MS. This is not particularly
surprising given that we are imaging the central regions of these clusters where
mass segregation has enhanced the binary populations. 
We should also note that some of these CMDs may contain significant white
dwarf sequences. Confirmation of these sequences will require star/galaxy
image classification, which will be presented in a future paper.

To fully appreciate the properties of these CMDs, it is important to examine
them within the context of existing ground-based photometry of these clusters. 

\noindent{\it NGC~5466:} The most recent CMDs of this cluster have 
been presented by Jeon et al.\ (2004), Rosenberg et al.\ (2000b), and 
Corwin, Carney, \& Nifong (1999). Because of its substantial population 
of blue stragglers, the majority of prior CMD studies have focused on these 
stars and their photometric variability. This cluster is of exceptionally 
low stellar density even near its center; it is probably because of this
that NGC~5466 has not been
a primary target of {\sl HST} studies. However, it is clear from Fig.\ 1 that our 
{\sl HST/ACS} CMD has revealed the principal sequences 
of NGC~5466 with significantly higher precision than previous 
observations.

\noindent{\it NGC~6779:} Hatzidimitriou et al.\ (2004) and Rosenberg et al.\
(2000b) have both published CCD-based CMDs for this cluster that reach
past the main sequence turnoff. The blue
HB and relatively metal-poor nature of the cluster were noted by both
studies.  Additionally, our CMD exhibits a population of blue straggler
stars as well as a better defined MSTO and unevolved MS suitable for
age determination.

\noindent {\it NGC~5053:} Structurally similar to NGC~5466 discussed above, 
this cluster also presents an exceptionally low stellar density, which is probably
the main reason it has not been previously observed by {\sl HST}. In addition, it 
also harbors a healthy population of blue straggler stars making it the focus
of numerous stellar brightness variability studies. Color-magnitude diagrams have
been presented by Rosenberg et al.\ (2000b), Sarajedini \& Milone
(1995), and Fahlman, Richer, \& Nemec (1991), among others.

\noindent{\it NGC~6144:} The sole published CMD of this cluster is
by Neely, Sarajedini, \& Martins (2000); their $BVI$ photometry reveals
a cluster with a predominantly blue HB and significant differential
reddening across its angular extent. This is because NGC~6144 is
viewed behind the $\rho$ Ophiuchi dust cloud approximately 40 arcmin 
northeast of the globular cluster NGC~6121 (M~4). 

\noindent{\it Palomar~2:} Located in a direction of extremely high 
foreground absorption and differential reddening toward the Galactic 
anti-center, the CMD of this cluster presented by Harris et al.\ (1997)
exhibits a significant amount of scatter. Nonetheless, these authors
were able to characterize the HB morphology of Pal 2 as being bi-modal
in color similar to that of NGC~1851, NGC~6229, and NGC~1261. 
As a result, they suggest that Pal~2 has a metallicity in the range 
{\rm [Fe/H]}$\sim$$-1.3 \pm 0.2$. The features seen by Harris et al.\
(1997) are also present in Fig.\ 5 with the exception of a
second MSTO feature blueward of the dominant MS at $F606W\sim23.5$ and 
$(F606W-F814W)\sim1.4$. We return to an analysis of this feature
and the metallicity of Pal~2 in Sec.\ 4.2. We note in passing that Pal 2 
has gained renewed prominence 
recently because of a suggestion by Majewski et al. (2004) that it could be 
associated with the Sagittarius dwarf spheroidal galaxy. 


\noindent{\it E~3:} The first CMD of this cluster was presented by van den Bergh,
Demers, \& Kunkel (1980) and later improved upon by McClure
et al.\ (1985). Both of these studies noted two intriguing characteristics of this cluster. 
First, the equal-mass
binary sequence in E~3 was among the most prominent ever observed to-date
among globular clusters. Second,
the cluster shows little or no indication of an HB population. Taken together,
along with the dramatic decrease in stellar density along the lower MS, 
these findings suggest the presence of significant mass segregation and/or 
mass loss in E~3.
Furthermore, these characteristics are consistent with the Galactic GC Pal~13 
which Siegel et al. (2001) argue is in the last stages of being destroyed.
The fact that E~3 does not possess a clear HB 
makes its distance and age determinations necessarily more
uncertain. The most recent CMD by Rosenberg et al.\ (2000a) corroborates
these results.

\noindent{\it Lyng\aa~7:} Thought to be an open cluster originally,
Lyng\aa~7 was studied photometrically by Ortolani, Bica, \& Barbuy
(1993), who showed that it is more likely to be a (thick) disk globular cluster.
Their CMD is heavily contaminated by non-cluster stars, but does
reveal a prominent core-helium burning red clump and a putative main
sequence turnoff that suggested an age significantly younger than other 
Galactic globular clusters at its metallicity. However, a reanalysis of the 
Ortolani et al.\ (1993) CMD by Sarajedini (2004) indicates an age close 
to that of 47 Tuc. Our CMD is also heavily contaminated by non-cluster
stars, but in the next section, we present a radially limited CMD which
shows the cluster sequences better.

\noindent{\it Palomar~1:} First photometered by Ortolani \& Rosino (1985)
and later by Borissova \& Spassova (1995), Pal~1 is similar to other
low-density globular clusters such as Pal~13 and E~3 in that  it exhibits
an RGB and HB that are both very poorly populated. The more recent
work of Rosenberg et al.\ (1998a; 1998b) has revealed a metallicity 
comparable with that of 47 Tuc, but an age that is some 4 to 5 Gyr 
{\it younger} than 47 Tuc. The possibility that Pal~1 has been misclassified
as a globular cluster cannot be ruled out. In fact, Crane et al. (2003) have 
shown that Pal~1 could be a member of the Monoceros stream
which has both open, transitional, and globular clusters as members 
(Frinchaboy et al. 2004), which is very interesting considering the 
ambiguity of Pal~1's classification.

\noindent{\it NGC~6366:} From the earliest CMD of this cluster by
Pike (1976) to the most recent by Alonso et al.\ (1997) and Rosenberg
et al. (2000b), it has been
recognized that NGC~6366 is a close twin of 47~Tuc. Both clusters have
predominantly red HBs, a metal abundance close to {\rm [Fe/H]}$\sim-0.7$ and 
a sparse RR Lyrae population, and, in fact, NGC 6366 is somewhat closer
to us than 47~Tuc. The main reason NGC~6366 has received much less
attention than 47~Tuc is because of its high extinction and significant 
differential reddening. 

\vskip 24pt
It is clear from the above discussion that many of the CMD features of
these clusters have already been revealed by ground-based observations,
especially those at magnitude levels above the MSTO.
What makes the {\sl HST} CMDs in Figs. 1 through 9 unique is that they
represent the deepest photometry for these clusters published to-date.
Most extend as much as seven magnitudes below the MSTO, in some cases,
revealing a sequence of unresolved binaries that has never before
been identified in these clusters. The binary populations of our target
clusters will be discussed in a  future paper.
The {\sl HST} CMDs presented herein are ideal for the determination of main sequence
fitting distances, which is the subject of the next section, and ages, which are
covered in Sec.\ 5. 

\section{Main Sequence Fitting}

To minimize systematic errors associated with our filter set, we have chosen to
perform main sequence fitting (MSF) in the ACS F606W and F814W filters. We have
selected three `standard' clusters from our Treasury database - M92, NGC~6752, and
47~Tuc - as comparison clusters. Our fiducial sequences for M92 and NGC~6752
are consistent with those presented by Brown et al. (2005) and with the 
ground-based data of Stetson (2000) in the $VI$ passbands.
However, the Brown et al.\ (2005) fiducial for 47~Tuc is significantly
bluer ($\sim$0.04 mag) than our Treasury data and the 
photometry of Stetson (2000). We will address these issues more fully in
a future paper, but, for the moment, it is important to point out that
the MSF performed herein makes use of photometry for the target and comparison
clusters that is internally consistent.


\subsection{Cluster Properties}

%
%
%
%

Table 2 lists the properties of the target and comparison clusters in the
present study. The metallicities on the Zinn \& West (1984)
and Carretta \& Gratton (1997) scales are taken from a number of
sources. For NGC~5466, 5053, 6366, 6752, M~92, and 47~Tuc, 
both {\rm [Fe/H]}$_{\rm ZW}$ and
{\rm [Fe/H]}$_{\rm CG}$ come from the globular cluster age-dating study of
De Angeli et al.\ (2005), while for Pal~1, these values are measured by
Rosenberg (1998b). The ZW and CG metallicities of NGC~6779 are
derived by Hatzidimitriou et al.\ (2004) as part of their $BVRI$ photometric study.  
For NGC~6144, Lyng\aa~7, and E~3, the ZW values come from
Neely et al.\ (2000), Tavarez \& Friel (1995), and Harris (1996), 
respectively, while the CG metallicities are calculated using the equation in 
Sec.\ 2.1 of De Angeli et al.\ (2005). The metallicity of Pal~2 is derived below
in Sec.\ 4.2.

Columns 4 and 5 of Table 2 give the reddening and absolute distance modulus 
of each cluster from the Harris (1996) catalog. These have generally been 
calculated assuming $M_v(HB)$ = 0.15 {\rm [Fe/H]} + 0.80 along with the mean 
apparent magnitude of the HB. This means that in cases where no HB is
apparent (e.g. Pal~1, E~3) or differential reddening makes the level of the HB
difficult to assess (Pal~2), the distances could be highly uncertain. Columns
6 and 7 are the reddenings and distance moduli derived via MSF in this paper.
In the case of the comparison clusters, these values are taken from 
the study of Carretta et al.\ (2000), wherein globular cluster fiducials are 
fit to local subdwarfs with HIPPARCOS parallaxes. 

The MS fits of the comparison cluster fiducials to the cluster CMDs are shown
in Figs. 10 through 17. For NGC~6144, because its metal abundance lies between
those of two standard clusters (M~92 and NGC~6752), we have fit both fiducials
to NGC~6144 and interpolated the results. In performing the MS fits, we have 
shifted the fiducials of the 
comparison clusters in magnitude and color to match the unevolved MS of the 
target clusters. In particular, we match the data and the fiducial at a point
$\sim$2 magnitudes below the MSTO of an old population which occurs at 
$M_{\rm F606W}\sim+4.0$. The shifts are then converted to reddenings and
distance moduli using the color excess and absorption equations for a G2 type
star given by Sirianni et al.\ (2005), namely $E(F606W - F814W) = 0.98 E(B-V)$ 
and  $A_{\rm F606W} = 2.85 E(F606W - F814W)$. The resultant reddenings and 
distance moduli were derived differentially from those of the comparison clusters. 
We estimate 1-$\sigma$ random errors of  
$\pm$0.01 mag and $\pm$0.05 mag, respectively, in the $E(B-V)$
and $(m-M)_0$ values.

It is instructive to consider our distance modulus results in light of the Harris (1996)
values.
Excluding E~3 and Pal~1, the values in Table 2 reveal a mean difference of 
$\langle$$\Delta$$(m-M)_0$$\rangle$ = 0.07 $\pm$ 0.05 between our distance
moduli and those of Harris (1996) in the sense (Us--Harris).  This is not surprising given 
that the MSF distances of our comparison clusters show a mean difference 
in the same sense of $\langle$$\Delta$$(m-M)_0$$\rangle$ = 0.15 $\pm$ 0.02 relative
to the Harris (1996) values.  Here and below, we use the small sample statistical
formulae of Keeping (1962) to calculate the standard error of the mean.

Returning to E~3 and Pal~1,
neither of these clusters exhibits a significant HB population in its CMD making 
the determination of distance especially challenging. The origin of the Harris (1996)
distance for E~3 is quoted as the McClure et al. (1985) work. However, that paper
makes no mention of a distance modulus for E~3, so it is
not immediately clear what this distance is based on. As a result, it is
difficult for us to reconcile the rather large difference of $\Delta$$(m-M)_0$ = 1.35
mag between our distance modulus and that of Harris (1996) for E~3. 
In the case of Pal~1, the Harris (1996) values are taken from Rosenberg et al. 
(1998a) who used MSF to 47 Tuc to estimate Pal~1's distance. The
precise value of the Pal~1 distance modulus from Rosenberg et al. (1998a) 
is given as $(m-M)_0 = 15.25 \pm 0.25$; adopting a total error of $\pm$0.1 mag
for our distance modulus determination ($\sigma_{\rm random}$$\sim$0.05 mag and
$\sigma_{\rm systematic}$$\sim$0.10 mag added in quadrature) leads to a difference of 
$\Delta$$(m-M)_0 = 0.51 \pm 0.27$ in the sense (Us -- Rosenberg), which is 
not statistically significant.

Turning now to a discussion of our derived reddenings, the mean
difference between us and Harris is 
$\langle$$\Delta$$E(B-V)$$\rangle$ = 0.044 $\pm$ 0.013 in the sense
(Us--Harris). This excludes
Pal~2, which has a very high and uncertain value. In contrast, this difference
turns out to be $\langle$$\Delta$$E(B-V)$$\rangle$ = 0.015 $\pm$ 0.007 for
the comparison clusters. Based on this, one might argue that our derived
reddening values are systematically too high. However, a different picture
emerges when we compare to the reddening maps of Schlegel, Finkbeiner,
\& Davis (1998). Here we find a mean difference of 
$\langle$$\Delta$$E(B-V)$$\rangle$ = --0.010 $\pm$ 0.029 for the target clusters
and $\langle$$\Delta$$E(B-V)$$\rangle$ = --0.015 $\pm$ 0.014 for
the comparison clusters suggesting
agreement with the Schlegel et al. maps. Note that we have excluded
Pal~2 as above from this calculation and NGC~6144 which is located
behind the $\rho$ Ophiuchi dust cloud. We note in passing that the 
Schlegel et al. (1998) maps yield a reddening of $E(B-V) = 0.91$ for
NGC~6144.

In addition to the distances and reddenings that are derived from the 
MSF procedure, we can also comment on the ages of the target
clusters relative to the comparison clusters, which have similar metallicities.
In particular, of the 9 clusters of interest (the case of Pal 2 is addressed in
the next subsection), only Pal~1 appears to
be significantly younger than its comparison cluster. There is also some indication
that E~3 could be younger than 47 Tuc. While earlier work showed
that Pal~1 is a relatively young cluster, there is no such expectation for
E~3, as McClure et al.\ (1985) actually quote an age of 18 Gyr for this cluster.
However, a more definitive statement about the age of E~3 will have to
wait for a reliable abundance determination for this cluster. Previous papers on
Lyng\aa~7 (Ortolani et al.\ 1993; Sarajedini 2004) suggested that it may be
younger than 47~Tuc by as much as $\sim$4 Gyr. In contrast, our CMD
unequivocally shows that the two are about the same age. We will return to
the subject of cluster ages in Sec.\ 5. We should note as well that Pal~1 and E~3 show a 
strong depletion of low-mass stars.  Even without quantitative luminosity functions,
their CMDs show a striking decline of star numbers along the lower
main sequence.  To fully address this question will require, among other
things, a more complete understanding of the photometric completeness
of our data. We will address this phenomenon in a future paper.

\subsection{The Special Case of Pal~2}

Of the clusters in this study, Pal~2 is the most problematical in
terms of applying the MSF technique. The CMD in Fig.\ 5 presents a main
sequence that is so broadened by differential reddening that MSF seems
intractable. However, there is an intriguing extension of the MS blueward and
brightward of the MSTO / SGB region, as shown in the left panel of Fig.\ 18.
Taking a cue from the photometric study of this cluster by Harris et al.\ (1997),
we hypothesize that this feature represents a minimally reddened population
in Pal~2 and proceed to investigate the spatial distribution of these stars.
The filled circles in the right panel of Fig.\ 18 are the stars within
the bounded region indicated in the left panel. The appearance of this figure
suggests the presence of a region of minimal reddening in the southeast 
quadrant of the cluster. The concentric circles represent zones approximately
centered on this region of Pal~2 for which CMDs are plotted in Fig.\ 19. The
exact center of these circles has been chosen to yield a CMD in the
innermost zone that best defines the cluster's principal sequences. The
solid line in Fig.\ 19 is the fiducial sequence of NGC~6752 (Brown et al.\ 2005)
shifted to match the main sequence of Pal 2 inside of 25$''$ from the center. 
The reddening and distance of Pal 2 listed in Table 2 are derived from
these shifts relative to NGC~6752. We estimate an error of $\pm$0.05 in
$E(B-V)$ and $\pm$0.10 mag in distance modulus.

Figure 19 corroborates our earlier hypothesis that the stars bounded by the
region in the left panel of Fig.\ 18 represent a population that is
minimally reddened. As we move further from the southeast quadrant of the cluster,
the reddening and its range increase so that the cluster sequences are redder
and show greater scatter. In light of this apparent trend in the differential reddening,
we have attempted to correct for its effects in the following manner. 
The first step involves the construction of a fiducial sequence for
the cluster.  Then each individual star yields a color residual,
taken along the reddening direction in the CMD.  From these residuals a
reddening map is made, by finding the median residual in each
256x256-pixel square of the image.  Then each star is corrected, along
its reddening line, by an amount that is interpolated from the 16x16
points of the reddening map.

The result of this procedure is shown in Fig.\ 20 along with the fiducial 
sequence of NGC~6752 shifted along the reddening vector to match the
principal sequence of Pal~2. The CMD of Pal~2 has tightened-up
considerably better-defining the HB and the so-called `RGB bump' that results
from a pause in the brightward evolution of RGB stars (Fusi Pecci 
et al.\ 1990; Sarajedini \& Forrester 1995; Ferraro et al.\ 1999). 

The luminosity of the RGB Bump depends on the cluster metallicity
and, to a lesser degree, the cluster age (Alves \& Sarajedini 1999). As such, 
given that Pal~2 is likely to be an old cluster,
we can use the magnitude difference between its HB and RGB Bump
to estimate its metal abundance.
The bottom panel of Fig.\ 20 illustrates the luminosity functions (LFs)
of the RGB [1.7$<$$(F606W-F814W)$$<$2.3, filled circles] and HB
[1.3$<$$(F606W-F814W)$$<$1.7, open circles], while the solid and
dashed lines are the Gaussian fits to these distributions over the range
20.0$<$$F606W$$<$21.6. These fits yield a difference
between the HB and Bump magnitudes of 
$\Delta$$F606W^{\rm Bump}_{\rm HB}$ = --0.30 $\pm$ 0.02. The error has been
calculated by the adding the standard errors of the mean HB and Bump magnitudes 
in quadrature. Assuming that
$\Delta$$F606W^{\rm Bump}_{\rm HB}$$\approx$$\Delta$$V^{\rm Bump}_{\rm HB}$,
the relations given in Table 6 of Ferraro et al.\ (1999) give metallicities of
{\rm [Fe/H]$_{\rm ZW} = -1.68 \pm 0.04$ and {\rm [Fe/H]$_{\rm CG} = -1.42 \pm 0.04$. 
The quote errors on these quantities represent only the random errors.
Thus, the metal abundance of Pal~2 is close to that of NGC~6752.
In addition, the close match between the fiducial of
NGC~6752 and the Pal~2 CMD suggests that the two
clusters have comparable ages and metallicities. 

\section{Isochrone Comparisons}

\subsection{Construction of Theoretical Isochrones}

Isochrones were generated from stellar evolution tracks produced by the 
Dartmouth Stellar Evolution Program (DSEP).  DSEP employs high temperature 
opacities from OPAL (Iglesias \& Rogers 1996), low temperature opacities from 
Ferguson et al.\ (2005), the detailed equation of state code FreeEOS (Irwin 2004)
 for low mass stars, and surface boundary conditions derived from PHOENIX 
 model atmospheres (Hauschildt et al.\ 1999a;1999b).  
 The physical inputs (e.g.  opacities, equation of state and convection
theory) used in the stellar interior models are consistent with the
physics used in the model atmosphere calculations.  A detailed discussion
of these new stellar models, color-Teff relations and isochrones is
presented in Dotter et al. (2007, in preparation).
 
A grid of stellar evolution tracks were computed for {\rm [Fe/H}]= --2.5, --2.0, --1.5, --1.0, --0.5, 
and 0.0 with [$\alpha$/Fe] = --0.2, 0.0, 0.2, 0.4, 0.6, and 0.8.  Opacities and surface 
boundary conditions were created for each specific composition listed.  Stellar masses 
between 0.1 and 1.5 ${\cal M}_{\odot}$ allow for isochrones with ages ranging from 3 to 15 Gyr  
that extend from $M_V \sim$ 14 to the tip of the RGB.

For comparison purposes, the isochrones were transformed using two different 
color-T$_{\rm eff}$ relations.  Both transformations include standard ground-based 
$B$, $V$, and $I$ magnitudes along with {\sl HST} filters F606W and F814W (both ACS-WFC and 
WFPC2).  The semi-empirical transformation uses the $B$, $V$, and $I$ magnitudes from 
VandenBerg \& Clem (2003) and the relevant equations to convert from $V$ and $I$ to 
F606W and F814W in Appendix D of Sirianni et al.\ (2005).  The synthetic 
color-T$_{\rm eff}$ transformation uses fluxes from the PHOENIX model atmospheres 
(also used for surface boundary conditions) along with the definitions of $B$, $V$, and $I$ 
from Bessell (1990) and {\sl HST} ACS/WFC and WFPC2 filters from Sirianni et al.\ (2005), 
all normalized to the Vega system.  The semi-empirical color transformation has the 
advantage of being constrained to fit observational data from globular clusters at low 
metallicities but does not explicitly account for the effects of $\alpha$-enhancement.  The 
synthetic color transformation, on the other hand, accounts for the influence of 
$\alpha$-enhancement but suffers from poor fits to the MSTO and subgiant regions that 
are typical of theoretical color transformations.

\subsection{Comparison to Cluster CMDs}

Given that the ZW and CG metallicities usually differ by less than $\sim$0.2 dex,
we adopt the latter abundance values in the isochrone fits. To select the 
appropriate ratio of the $\alpha$-elements to iron, we assume
that [$\alpha$/Fe] = +0.4 for {\rm [Fe/H]} $< -1.0$ and [$\alpha$/Fe] = +0.2 for 
{\rm [Fe/H]} $\sim  -0.7$. These [$\alpha$/Fe] values reflect the general trend observed 
in thick disk and halo stars (e.g. Origlia \& Rich 2004; Boesgaard et al. 2005; 
Mel{\'e}ndez et al. 2006; Reddy et al. 2006). We use the
MSF distance and reddening in Table 2 as a starting point; then we vary the
distance and reddening
until a satisfactory correspondence is achieved between the MS of
the cluster and the isochrone making sure that the point
$\sim$2 magnitudes below an old MSTO (i.e. $M_{\rm F606W}\sim+6.0$) matches, similar
to the MSF technique performed earlier. The results are shown in Figs. 21 through
28. Pal~2 is excluded because the MS is not adequately defined to allow a
comparison of the isochrone to the cluster's unevolved MS. However, as
noted above, its age is likely to be close to that of NGC~6752.

Examination of the isochrone fits reveals the following. The fit to the
unevolved MS stars is relatively good for the metal-poor clusters and degrades
for the empirical color-T$_{\rm eff}$ transformation as metallicity increases. 
In this case, the models tend to be redder than the observational data. 
In contrast, the synthetic transformation provides a better 
correspondence with the photometry of the MS.  The poor match to the lower 
main sequence could be due to either difficulities with the effective 
temperatures predicted by the models, or with the color-temperature 
relations.  We are currently investigating this issue.

As expected from the MSF results, the isochrone fits of NGC~5466, 6779, 
5053, 6144, Lyng\aa~7, and NGC~6366 reveal ages in the range of 12 to 14 Gyr. 
In contrast, the age of E~3 appears to be about 2 Gyr younger than these clusters
while the age of Pal~1 could be as much as 8 Gyr younger. The typical
error on these ages is between 1 and 2 Gyr primarily because they depend
on the uncertain distances and reddenings of the clusters. In a future paper, we
will address the issues of relative and absolute GC 
ages using more robust techniques. We emphasize that, as
noted in Sec.\ 4, the age of E~3 is likely to be especially uncertain until a more
reliable metallicity is determined for it. 

\section{Summary and Conclusions}

We have presented the first {\sl HST}-based CMDs for the Galactic globular 
clusters NGC~5446, 6779, 5053, 6144, Pal~2, E~3, Lyng\aa~7, Pal~1, and
NGC~6366. The CMDs extend reliably from the horizontal branch to as much as
seven magnitudes fainter than the main sequence turnoff. Features revealed
in these diagrams include the HB morphology, the presence of blue straggler
stars, unresolved-binary populations parallel to the main sequence, and 
possibly white dwarf populations. 

Using fiducial sequences for three standard clusters (M~92, NGC~6752, and 47~Tuc) 
with well-known metallicities and distance moduli, we perform main sequence fitting
on the target clusters in order to obtain estimates of their distances and reddenings. 
These comparisons along with fitting the cluster main sequences to theoretical
isochrones provide estimates of the clusters' ages. We find that only E3 and Pal 1
are significantly younger than the standard cluster to which they are compared.
E3 is $\sim$2 Gyr younger than 47 Tuc and Pal 1 could be as much as 8 Gyr
younger than 47 Tuc.
The ages of the remaining clusters are consistent with the standard clusters
at their metallicities.

\acknowledgments

Support for this work (proposal number 
GO-10775) was provided by NASA through a 
grant from the Space Telescope Science Institute 
which is operated by the Association of Universities 
for Research in Astronomy, Incorporated, under NASA contract NAS5-26555.

\clearpage

\begin{deluxetable}{ccccccccc}
\tablecaption{Observing Log}
\tablewidth{0pt}
\tabletypesize{\scriptsize}
\tablehead{
   \colhead{Cluster}
  &\colhead{$\alpha_{\rm 2000}$}
  &\colhead{$\delta_{\rm 2000}$}
  &\colhead{$l$}
  &\colhead{$b$}
  &\colhead{Dataset}
  &\colhead{UT Date}
  &\colhead{Filter}
  &\colhead{Exp Time}\\
     \colhead{}
  &\colhead{(h m s)}
  &\colhead{($^{\circ}$ $'$ $"$)}
  &\colhead{($^{\circ}$)}
  &\colhead{($^{\circ}$)}
  &\colhead{}
  &\colhead{}
  &\colhead{}
  &\colhead{}
}
\startdata
NGC~5466 & 14 05 27.3 & +28 32 04  & 42.15 & 73.59   &  J9L903  & 2006-04-12 &    F606W    &  1 x 30s,     5 x 340s  \\
                     &                      &                     &             &              &                  &                        &    F814W    & 1 x 30s,     5 x 350s  \\
                     &                      &                      &            &              &                 &                        &                     &                                       \\
NGC~6779 & 19 16 35.5 &  +30 11 05 & 62.66 &   8.34   &  J9L905  & 2006-05-11  &    F606W    &  1 x 20s,   5 x 340s   \\
      (M56)     &                      &                      &            &              &                  &                        &    F814W    &  1 x 20s,   5 x 350s   \\
                     &                      &                      &            &              &                  &                        &                     &                                           \\
NGC~5053 &13 16 27.0  &+17 41 53   & 335.69 & 78.94 & J9L902   & 2006-03-06 &    F606W    & 1 x 30s,     5 x 340s \\
                     &                      &                     &               &             &                 &                        &    F814W    & 1 x 30s,     5 x 350s \\
                     &                      &                     &               &             &                 &                        &                     &                                 \\
NGC~6144& 16 27 14.1  & --26 01 29  & 351.93 & 15.70  &  J9L943  & 2006-04-15 &   F606W     &  1 x 25s, 5 x 340s   \\
                     &                      &                     &                &             &                  &                        &   F814W     &  1 x 25s, 5 x 350s   \\
                     &                      &                     &                &             &                  &                        &                     &                                           \\
Palomar~2 &04 46 05.9  & +31 22 51 & 170.53   & --9.07  &  J9L908  & 2006-08-08 &   F606W    &   5 x 380s  \\
                     &                     &                     &                &              &                  &                       &    F814W    &   5 x 380s  \\
                     &                     &                     &                &              &                  &                       &                     &                    \\
E~3             & 09 20 59.3  & --77 16 57 & 292.27  & --19.02 &  J9L906   & 2006-04-15 &   F606W     & 1 x 5s, 4 x 100s  \\
                     &                     &                    &                &               &                   &                       &  F814W     &  1 x 5s, 4 x 100s  \\
                     &                     &                    &                &               &                   &                       &                     &                                           \\
Lyng\aa~7  &16 11 03.0  & --55 18 52 &  328.77 &  --2.79  &  J9L904   & 2006-04-07 &  F606W    & 1 x 35s, 5 x 360s   \\
                     &                     &                     &                &               &                   &                       &  F814W    & 1 x 35s, 5 x 360s   \\
                     &                     &                     &                &               &                   &                       &                     &                                           \\
Palomar~1 &03 33 23.0  & +79 34 50 &130.07   & 19.03    &  J9L901  & 2006-03-17  & F606W     &  1 x 15s, 5 x 390s         \\
                     &                     &                     &                &               &                  &                        & F814W     & 1 x 15s,  5 x 390s   \\
                     &                    &                      &                &               &                  &                        &                     &                                           \\
NGC~6366 & 17 27 44.3 & --05 04 36 &  18.41   & 16.04    &   J9L907 & 2006-03-30 &  F606W    &  1 x 10s, 4 x 140s  \\
                     &                     &                      &               &                &                 &                        &  F814W     & 1 x 10s, 4 x 140s  \\
\enddata
\end{deluxetable}

\clearpage

\begin{deluxetable}{lcccccc}
\tablecaption{Cluster Parameters}
\tablewidth{0pt}
\tabletypesize{\scriptsize}
\tablehead{
   \colhead{Cluster}
  &\colhead{$[Fe/H]_{\rm ZW}$}
  &\colhead{$[Fe/H]_{\rm CG}$}
  &\colhead{$E(B-V)$\tablenotemark{a}}
  &\colhead{$(m-M)_0$\tablenotemark{a}}
  &\colhead{$E(B-V)$\tablenotemark{b}}
  &\colhead{$(m-M)_0$\tablenotemark{b}}
}
\startdata
                                     &               &  &  Target Clusters &               &           &     \\
                                     &               &                  &               &               &           &     \\
NGC~5466                 &  --2.22   &    --2.20   &   0.00    &   16.00  &  0.02 & 16.03 \\
NGC~6779  (M~56)     &  --2.20   &    --2.00   &   0.20    &  15.03  &  0.26 &  15.08 \\
NGC~5053                 &  --2.10   &    --1.98   &   0.04    &   16.07  &  0.01 &  16.22 \\
NGC~6144                 &  --1.81   &    --1.56   &   0.36     &  14.64  &  0.45 &  14.61  \\
Palomar~2                 &  --1.68    &    --1.42   &   1.24     &   17.21  & 0.94 &  17.13 \\
E~3                               &  --0.80   &     --0.83  &  0.30      &   13.19 & 0.34 & 14.54  \\
Lyng\aa~7                &  --0.62   &    --0.64   &  0.73      &   14.28  & 0.78 & 14.55     \\
Palomar~1                 &  --0.6      &   --0.7      &  0.15     &    15.19  & 0.23 & 15.76   \\
NGC~6366                &  --0.58    &  --0.73    &  0.71      &    12.77  & 0.75 & 12.87    \\
                                    &               &                  &               &                  &            &     \\
\hline
                                    &               &                  &               &                 &            &     \\
                                    &               &  & Comparison Clusters  &       &            &     \\
                                    &               &                  &               &                &            &     \\
NGC~6341 (M~92)     &  --2.24   &    --2.16   &   0.02    &   14.58   & 0.025  & 14.66    \\
NGC~6752                 &  --1.54   &    --1.24   &   0.04    &   13.01  &  0.035  & 13.23    \\
NGC~104 (47~Tuc)   &  --0.71   &    --0.78   &   0.04    &   13.25   &  0.055 &  13.40    \\
\enddata
\tablenotetext{a}{Data from Harris (1996). Absolute distance modulus has been
computed assuming $A_V = 3.1E(B-V)$.}
\tablenotetext{b}{Results from main sequence fitting.}
\end{deluxetable}

\clearpage

\begin{figure}
\caption{The color-magnitude diagram for NGC 5466 in the VEGAmag system. The
plus symbols represent stars that are affected
by at least one saturated pixel in either or both of the F606W and F814W images. 
This diagram contains 21,449 stars and extends to approximately 12\% of the
tidal radius of 34$'$ (Harris 1996).}
\end{figure}


\begin{figure}
\caption{Same as Fig.\ 1 except that the CMD of NGC 6779 (M56) is shown 
containing 61,056 stars and extending to about 50\% of the tidal radius
of 8.5$'$ (Harris 1996).}
\end{figure}


\begin{figure}
\caption{Same as Fig.\ 1 except that the CMD of NGC 5053 is shown
containing 15,618 stars and extending to about 30\% of the tidal radius
of 14$'$ (Harris 1996).}
\end{figure}


\begin{figure}
\caption{Same as Fig.\ 1 except that the CMD of NGC 6144 is shown 
containing 19,442 stars and extending to about 13\% of the tidal radius
of 33$'$ (Harris 1996).}
\end{figure}


\begin{figure}
\caption{Same as Fig.\ 1 except that the CMD of Palomar 2 is shown
containing 43,242 stars and extending to about 60\% of the tidal radius
of 6.8$'$ (Harris 1996). The arrow is the reddening vector for 
$E(F606W - F814W) = 0.3$.}
\end{figure}


\begin{figure}
\caption{Same as Fig.\ 1 except that the CMD of E3 is shown
containing 852 stars and extending to about 40\% of the tidal radius
of 11$'$ (Harris 1996).}
\end{figure}


\begin{figure}
\caption{Same as Fig.\ 1 except that the CMD of Lyng\aa 7  is shown
containing 32,226 stars. The tidal radius for Lyng\aa 7 is not given in
the Harris (1996) compilation. The arrow is the reddening vector for 
$E(F606W - F814W) = 0.3$.}
\end{figure}


\begin{figure}
\caption{Same as Fig.\ 1 except that the CMD of Palomar 1 is shown 
containing 808 stars  and extending to about 47\% of the tidal radius
of 9.0$'$ (Harris 1996).}
\end{figure}


\begin{figure}
\caption{Same as Fig.\ 1 except that the CMD of NGC 6366 is shown 
containing 5,503 stars  and extending to about 27\% of the tidal radius
of 15$'$ (Harris 1996).}
\end{figure}


\begin{figure}
\caption{The result of fitting the fiducial sequence of M92 
to the main sequence of NGC 5466. The small open squares are M92 HB stars. 
The plus symbols represent stars that are affected
by at least one saturated pixel in either or both of the F606W and F814W images.}
\end{figure}

\clearpage

\begin{figure}
\caption{Same as Fig.\ 10 except that the CMD of NGC 6779 (M56) is shown. }
\end{figure}


\begin{figure}
\caption{Same as Fig.\ 10 except that the CMD of NGC 5053 is shown. }
\end{figure}


\begin{figure}
\caption{The result of fitting the fiducial sequences of M92 (left panel) and NGC 6752
(right panel) to the main sequence of NGC 6144 . The small
open squares are the HB stars of M92 (left) and NGC 6752 (right). The plus symbols 
represent stars that are affected
by at least one saturated pixel in either or both of the F606W and F814W images.}
\end{figure}


\begin{figure}
\caption{The result of fitting our fiducial sequence of 47 Tuc 
to the main sequence of E3. The small open squares the HB stars in 47 Tuc.
The plus symbols 
represent stars that are affected
by at least one saturated pixel in either or both of the F606W and F814W images.}
\end{figure}


\begin{figure}
\caption{Same as Fig.\ 14 except that the CMD of Lyng\aa 7 is shown for
stars less than 15$''$ from the cluster center. }
\end{figure}


\begin{figure}
\caption{Same as Fig.\ 14 except that the CMD of Pal 1 is shown.  }
\end{figure}


\begin{figure}
\caption{Same as Fig.\ 14 except that the CMD of NGC 6366 is shown. }
\end{figure}


\begin{figure}
\epsscale{1.2}
\caption{The left panel shows the CMD of Pal 2 with the main sequence 'extension'
indicated by a box-like region. The right panel shows the location of stars (filled circles)
in this box relative to other stars in the cluster field. The circles are drawn at 
radii of 25$''$, 50$''$, 100$''$, and 150$''$ and centered at (1800,2400) pixels. }
\end{figure}


\begin{figure}
\epsscale{0.8}
\caption{The radial CMDs of Pal 2 centered on (1800,2400) pixels (see Fig.\ 11).
The fiducial and HB stars are those of NGC 6752 shifted to the distance and
reddening of the data in the inner region (upper left panel). }
\end{figure}


\begin{figure}
\epsscale{1.0}
\caption{The upper panel shows the differential-reddening-corrected CMD
for Pal 2 along with the fiducial sequence of NGC~6752 shifted along the 
reddening vector to match the principal sequences of Pal 2.  The lower
panel displays the RGB (filled circles) and HB (open circles) LFs of Pal 2 
along with the corresponding Gaussian fits (solid and dashed lines).}
\end{figure}


\begin{figure}
\caption{The CMD of NGC 5466 along with theoretical isochrones based on two
different color-$T_{\rm eff}$ transformations (see text). }
\end{figure}


\begin{figure}
\caption{Same as Fig.\ 21 except that the CMD of NGC 6779 (M56) is plotted. }
\end{figure}


\begin{figure}
\epsscale{1.0}
\caption{Same as Fig.\ 21 except that the CMD of NGC 5053 is plotted. }
\end{figure}

\begin{figure}
\caption{Same as Fig.\ 21 except that the CMD of NGC 6144 is plotted. }
\end{figure}


\begin{figure}
\caption{Same as Fig.\ 21 except that the CMD of E3 is plotted. }
\end{figure}


\begin{figure}
\caption{Same as Fig.\ 21 except that the CMD of Lyng\aa 7 is plotted for stars
inside of 15$''$ from the cluster center. }
\end{figure}


\begin{figure}
\epsscale{1.0}
\caption{Same as Fig.\ 21 except that the CMD of Palomar 1 is plotted. }
\end{figure}


\begin{figure}
\epsscale{1.0}
\caption{Same as Fig.\ 21 except that the CMD of NGC 6366 is plotted. }
\end{figure}

\end{document}